\documentclass[aps,preprint,showkeys]{revtex4-1}
\usepackage{graphicx}
\usepackage{amsmath}
\usepackage{makeidx}
\usepackage{amsfonts}
\usepackage{amssymb}
\usepackage{mathtools}
\usepackage{xcolor}

\begin{document}

\title{Noise tolerance via reinforcement: Learning a reinforced quantum dynamics}
\date{\today}

\author{Abolfazl Ramezanpour}
\email{aramezanpour@gmail.com}
\affiliation{Department of Physics, College of Science, Shiraz University, Shiraz 71454, Iran}
\affiliation{Leiden Academic Centre for Drug Research, Faculty of Mathematics and Natural Sciences, Leiden University, PO Box 9500-2300 RA Leiden, The Netherlands}

\begin{abstract}
The performance of quantum simulations heavily depends on the efficiency of noise mitigation techniques and error correction algorithms. Reinforcement has emerged as a powerful strategy to enhance the efficiency of learning and optimization algorithms. In this study, we demonstrate that a reinforced quantum dynamics can exhibit significant robustness against interactions with a noisy environment. We study a quantum annealing process where, through reinforcement, the system is encouraged to maintain its current state or follow a noise-free evolution. A learning algorithm is employed to derive a concise approximation of this reinforced dynamics, reducing the total evolution time and, consequently, the system’s exposure to noisy interactions. This also avoids the complexities associated with implementing quantum feedback in such reinforcement algorithms. The efficacy of our method is demonstrated through numerical simulations of reinforced quantum annealing with one- and two-qubit systems under Pauli noise.    
\end{abstract}

\maketitle

\section{Introduction}\label{S0}
Quantum systems exhibit exceptional sensitivity to systematic noise and environmental interactions, making noise mitigation essential for practical quantum simulations and algorithms \cite{Knill-nat-2005}. Fault-tolerant quantum computation—leveraging topological quantum states and advanced error-correcting codes—provides partial solutions \cite{Bharti-rmp-2022}. However, achieving quantum advantage ultimately demands large-scale systems with numerous logical (and consequently physical) qubits \cite{Harrow-nat-2017,Boixo-np-2018}. This fundamental requirement underscores the critical challenge of operating many physical qubits in a noisy environment \cite{Kim-nat-2023,P-acm-2025}.

In principle, physical implementations of quantum algorithms are susceptible to two nearly unavoidable types of errors: (1) errors in preparing input and measuring output states, and (2) errors arising during system dynamics. Various error mitigation schemes, such as repetition and error modeling, have been developed to address the former errors \cite{Smith-savd-2021,Bravyi-pra-2021,Hicks-pra-2022,Elben-nrp-2023}. The latter category includes Hamiltonian uncertainties (coherent noise) and environmental interactions (incoherent noise). Systematic and coherent errors are particularly computationally demanding to correct \cite{Green-qst-2017,Cai-qi-2020,Zhang-prap-2022,Proctor-prl-2022,Van-np-2023}. On the other hand, specialized error-correcting codes exist for specific noise environments. These may involve techniques such as noise strength extrapolation or noise estimation \cite{Li-prx-2017,He-pra-2020,Sun-pap-2021,Urbanek-prl-2021,Yi-prl-2024}. Additionally, artificial neural networks and machine learning methods have shown promise in predicting and correcting such errors \cite{Kim-iee-2020,Strikis-prx-2021,Liao-nml-2024,Wang-repp-2024,Bausch-nat-2024}. In this study, we propose a noise mitigation method designed to deal with incoherent noise and reduce engineering errors.

Numerous learning and optimization algorithms benefit from reinforcements (or feedback mechanisms) which exploit the information obtained from the system to enhance the performances \cite{SB-book-1998,Alf-prl-2006,baldassi-pnas-2007,QC-prep-2013,QR-pra-2017,RQA-scirep-2020,Lin-pra-2020,Moos-ml-2022}. Reinforcement is used to guide the system dynamics through a complex search space while reshaping the energy landscape to favor the solution subspace. Notably, reinforcement can reduce the disturbing effects of exponentially small energy gaps near quantum phase transitions in quantum annealing simulations \cite{QR-pra-2022}. The inherent nature of such feedback mechanisms suggests that they should also confer robustness against noise in a reinforced quantum dynamics - a phenomenon we explicitly demonstrate in a quantum search problem \cite{grover-prl-1997,Roland-pra-2002,fdsearch-prl-2005,fdsearch-pra-2017,search-pra-2020}. This is particularly significant as quantum search algorithms exhibit notable noise sensitivity, with only limited mitigation methods currently available \cite{Pablo-pra-1999,Shapira-pra-2003,Salas-epj-2008,Pan-qi-2023,Leng-arx-2023,Pati-jpa-2024,Leng-prr-2025}. 

We also develop a learning algorithm, within the teacher-student framework, to get around the physical implementation costs of reinforcements in a teacher model. We construct a student model with a simpler and shorter quantum dynamics than the teacher to reproduce approximately the outputs of a reinforced dynamics. This further reduces noise disturbances while maintaining the benefits of reinforcement in the original (teacher) model.

\begin{figure}
	\includegraphics[width=12cm]{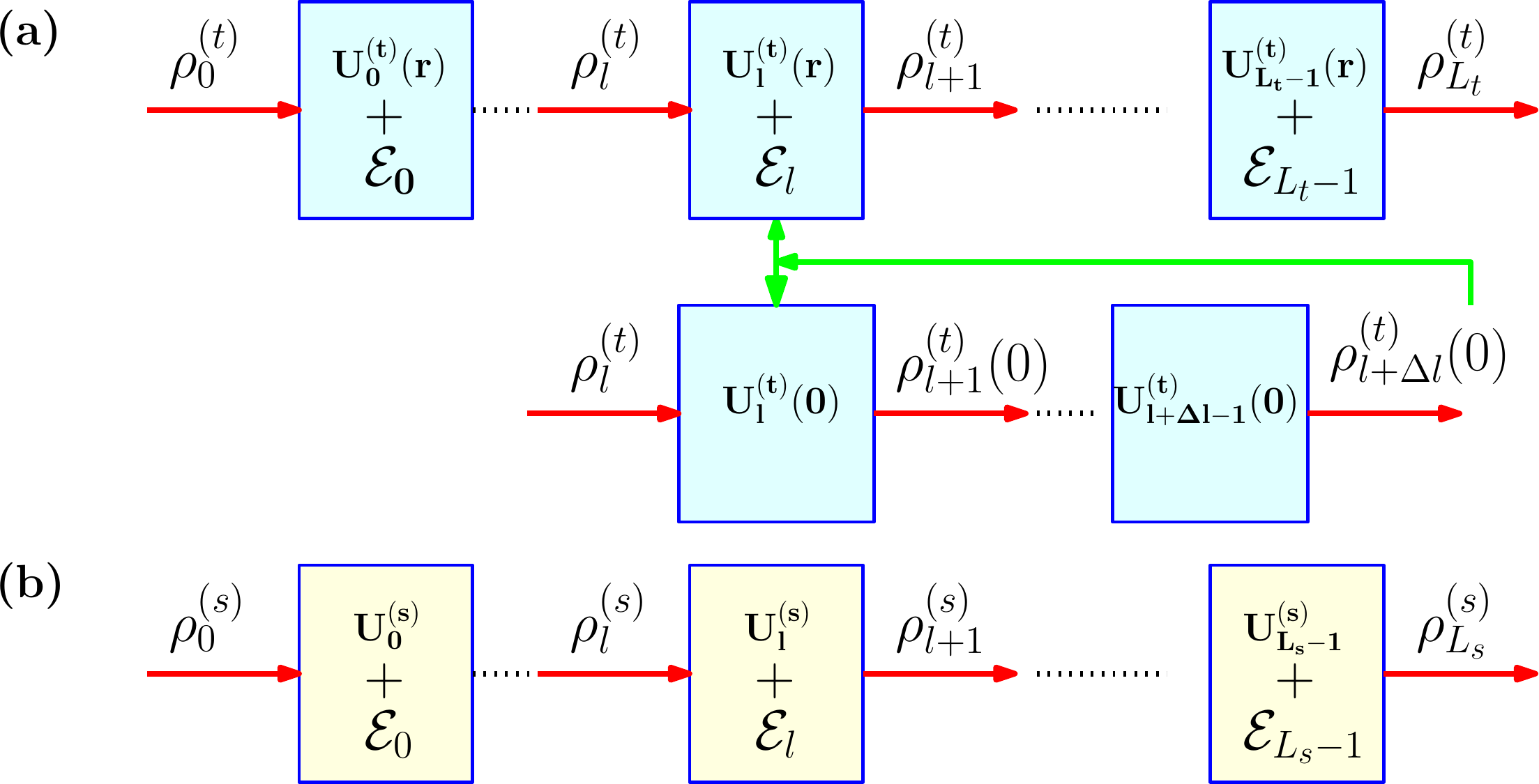} 
	\caption{Dynamical models of the teacher and student under noise. (a) The teacher’s dynamics is reinforced at each layer $l$ by steering toward $\rho_{l+\Delta l}^{(t)}(0)$ (ideal noise-free state). (b) The student’s unitaries are trained to exclusively replicate the teacher’s noise-free reinforced dynamics.}\label{fig-M000}
\end{figure}

\section{Teacher model}\label{S1}
Consider a teacher dynamical model of $L_t$ consecutive layers of unitary evolutions with $U_l^{(t)}$ and quantum maps $\mathcal{E}_l$ for $l=0\cdots,L_t-1$, see Fig. \ref{fig-M000}. The latter is used to represent a noisy channel (environment). The quantum states in each layer of the model live in a Hilbert space of dimension $d$.
For an input state $\rho_0$ to the teacher, we obtain the sequence of states $\rho_{l+1}=\mathcal{E}_l(U_{l}^{(t)}\rho_l U_{l}^{(t)\dagger})$ up to the output state $\rho_{L_t}$. More precisely, the quantum states of the teacher evolve as follows
\begin{align}\label{rho-l}
\rho_{l+1} &=(1-\epsilon_l)U_{l}^{(t)}\rho_l U_{l}^{(t)\dagger}+\epsilon_l\sum_{\alpha}K_{l,\alpha}(U_{l}^{(t)}\rho_l U_{l}^{(t)\dagger})K_{l,\alpha}^{\dagger}.
\end{align}
The Kraus operators $K_{l,\alpha}$ in each layer satisfy $\sum_{\alpha} K_{l,\alpha}K_{l,\alpha}^{\dagger}=\mathbb{I}$. The associated evolution $\sum_{\alpha}K_{l,\alpha}\rho_l K_{l,\alpha}^{\dagger}$ models the effect of noise on the system, with the parameter $0\le \epsilon_l\le 1$ controlling the noise strength. For simplicity, we assume that the strength of noise is independent of layer index, $\epsilon_l=\epsilon/L_t$. 

In this study, the unitary part of the evolution $U_{l}^{(t)}(r) =e^{-\hat{i}H_l^{(t)}(r)}$ is used to represent a reinforced quantum annealing process \cite{QR-pra-2017}. In the following we set $\hbar=1$. The reinforced Hamiltonian at layer $l$ is 
\begin{align}
H_l^{(t)}(r) &=(1-t_l)H_i+t_lH_f+r_lH_r(\rho_l),\\
H_r(\rho_l) &=-\ln \rho_{l+\Delta l}(0),
\end{align}
depending on the initial and final Hamiltonians $H_i, H_f$ with ground states $|\psi_i \rangle$ and $|\psi_f \rangle$, respectively. 
The evolution time $t_l\in (0,1)$ is a monotonically increasing function of layer index $l$. The last term $H_r$ reinforces the evolution using the quantum state of the system after $\Delta l$ layers of unitary evolutions in absence of noise ($\epsilon=0$) and reinforcement ($r=0$), that is 
\begin{align}
\rho_{l+1}(0) &=U_{l}^{(t)}(0)\rho_lU_{l}^{(t)\dagger}(0),\\ 
\rho_{l+2}(0) &=U_{l+1}^{(t)}(0)\rho_{l+1}(0)U_{l+1}^{(t)\dagger}(0), \dots 
\end{align}
up to $\rho_{l+\Delta l}(0)$. Note that the process begins with the current state at layer $l$, i.e., $\rho_{l}(0)=\rho_l$, see Fig. \ref{fig-M000}. Encouraging the system to maintain its current state ($\Delta l = 0$) or follow a noise-free trajectory ($\Delta l > 0$) is expected to improve noise robustness. For simplicity, we assume the reinforcement parameter is the same for all layers, $r_l = r$.
The success probability of the teacher with input state $\rho_0=|\psi_i \rangle\langle \psi_i|$ is the probability of finding the ground state of $H_f$ at the end of the dynamics, $P_{success}^{(t)}=\mathrm{Tr}(\rho_{L_t}|\psi_f \rangle\langle \psi_f|)$. 

\begin{figure}
	\includegraphics[width=16cm]{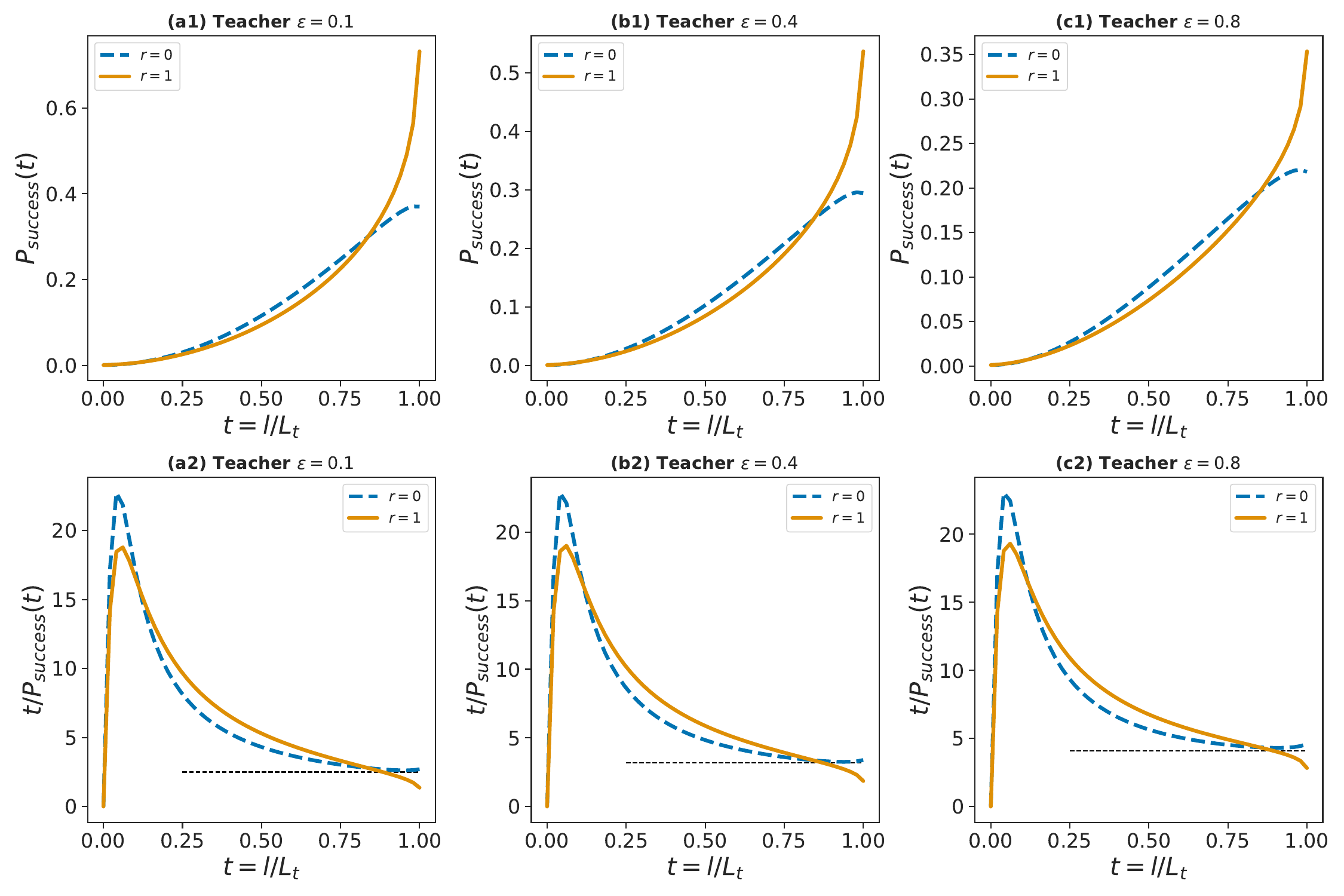} 
	\caption{Performance of the teacher model with $N=10$ qubits under Pauli noise channels. ((a1), (b1), (c1)) Success probability of the quantum annealing process, comparing cases with ($r=1$) and without ($r=0$) reinforcement.	((a2), (b2), (c2)) The scale of running time for the same data shown in the upper panels, as defined in Ref. \cite{Leng-prr-2025}. The teacher model consists of $L_t=50$ layers. The results are averaged over $100$ random and independent realizations of single-qubit Pauli noise channels. Statistical errors in success probabilities are less than $6\times 10^{-4}$.}\label{fig-M00}
\end{figure}

Figure \ref{fig-M00} illustrates the performances with $N=10$ qubits and $L_t=50$ evolution layers under Pauil noise channels.
We employ a quantum annealing process to represent a quantum search problem. The initial and final Hamiltonians are given by $H_i=\mathbb{I}-|\psi_i \rangle\langle \psi_i|$ and $H_f=\mathbb{I}-|\psi_f \rangle\langle \psi_f|$, where $|\psi_i \rangle=\frac{1}{\sqrt{2^N}}\sum_{\boldsymbol\sigma \in \{+,-\}^N}|\boldsymbol\sigma \rangle$ is a uniform superposition over all basis states and the target state is $|\psi_f \rangle=|+\rangle^{\otimes N}$. The system's current state $\rho_l$ is used for reinforcement (i.e., $\Delta l=0$), as a baseline reference. At each layer $l$ the noise is modeled by Pauli operators of weight one $K_{l,i\mu}=\sqrt{\frac{p_{l\mu}}{N}}\sigma_{\mu=x,y,z}^i$ acting on qubit $i$. The probabilities $p_{l\mu}\in (0,1)$ are uniformly distributed random variables which satisfy the normalization constraint $\sum_{\mu}p_{l\mu}=1$. We use the optimal annealing schedule provided by the Grover algorithm \cite{search-pra-2020}
\begin{align}\label{grover}
	t_l^* &=\frac{1}{2}\left[1-\sqrt{\frac{P_0}{1-P_0}}\tan\left((1-2\frac{l}{L_t-1})\alpha\right)\right],\\
	\alpha &=\arctan\left(\sqrt{\frac{1-P_0}{P_0}}\right),
\end{align}
which guarantees a quadratic speedup over classical search. This allows us to isolate the effects of reinforcement and noise on an otherwise optimal quantum search protocol.

Figure \ref{fig-M00} (upper panels) shows the success probability $P_{success}$ as a function of the number of layers for different noise strengths $\epsilon$. For comparison, we also include the results without reinforcement ($r=0$). The figure demonstrates that even with $\Delta l=0$, reinforcement ($r=1$) significantly improves $P_{success}$. Here, $P_{success}$ is evaluated at the end of the process ($l=L_t-1$). Alternatively, one could consider $P_{success}(l^*)$, where $l^*$ minimizes the scale of running time $l/P_{success}(l)$; this noise-tolerant approach leverages success-probability prediction to reduce the algorithm's runtime in noisy conditions \cite{Leng-prr-2025}. Figure \ref{fig-M00} (lower panels) displays the scale of running time for the same dataset. As expected, without reinforcement, the optimal $l^*$ shifts to smaller values as $\epsilon$ increases. Nevertheless, reinforcement consistently yields a better performance at the end of the annealing process.

The reinforced dynamics described above requires knowledge of the system's quantum state $\rho_l$. In principle, this state can be estimated via a complete set of collective weak measurements \cite{DK-pra-1999,lloyd-pra-2000,Qestimation-prl-2006,Qcontrol-book}. Given enough copies, the perturbations to the states in such weak measurements can be made arbitrarily small. To circumvent these practical challenges (e.g., resource constraints), we introduce a student model that replicates approximately the teacher dynamics by using ordinary (unreinforced) unitary evolutions.

\section{Student model}\label{S2}
We consider a student dynamical model consisting of $L_s$ unitary evolutions $U_l^{(s)}=e^{-\hat{i}H_l^{(s)}}$ in presence of noise. The to be learned Hamiltonians $H_l^{(s)}(\boldsymbol\theta_l)$ depend on parameters $\boldsymbol\theta_l$ for $l=0\cdots,L_s-1$. The student Hamiltonian for a system of $N$ qubits is limited to one- and two-spin interactions
\begin{align}
	H_l^{(s)}=\sum_{i=1}^N\sum_{\mu=x,y,z}\theta_{l,i\mu}\sigma_{\mu}^i+\sum_{i<j}\sum_{\mu,\mu'}\theta_{l,i\mu j\mu'}\sigma_{\mu}^i \sigma_{\mu'}^j,
\end{align}
with Pauli matrices $\sigma_{\mu=x,y,z}^i$.
The quantum states $\rho_{l}$ in the student model are defined in a Hilbert space of dimension $d$ as the teacher model.

\begin{figure}
	\includegraphics[width=12cm]{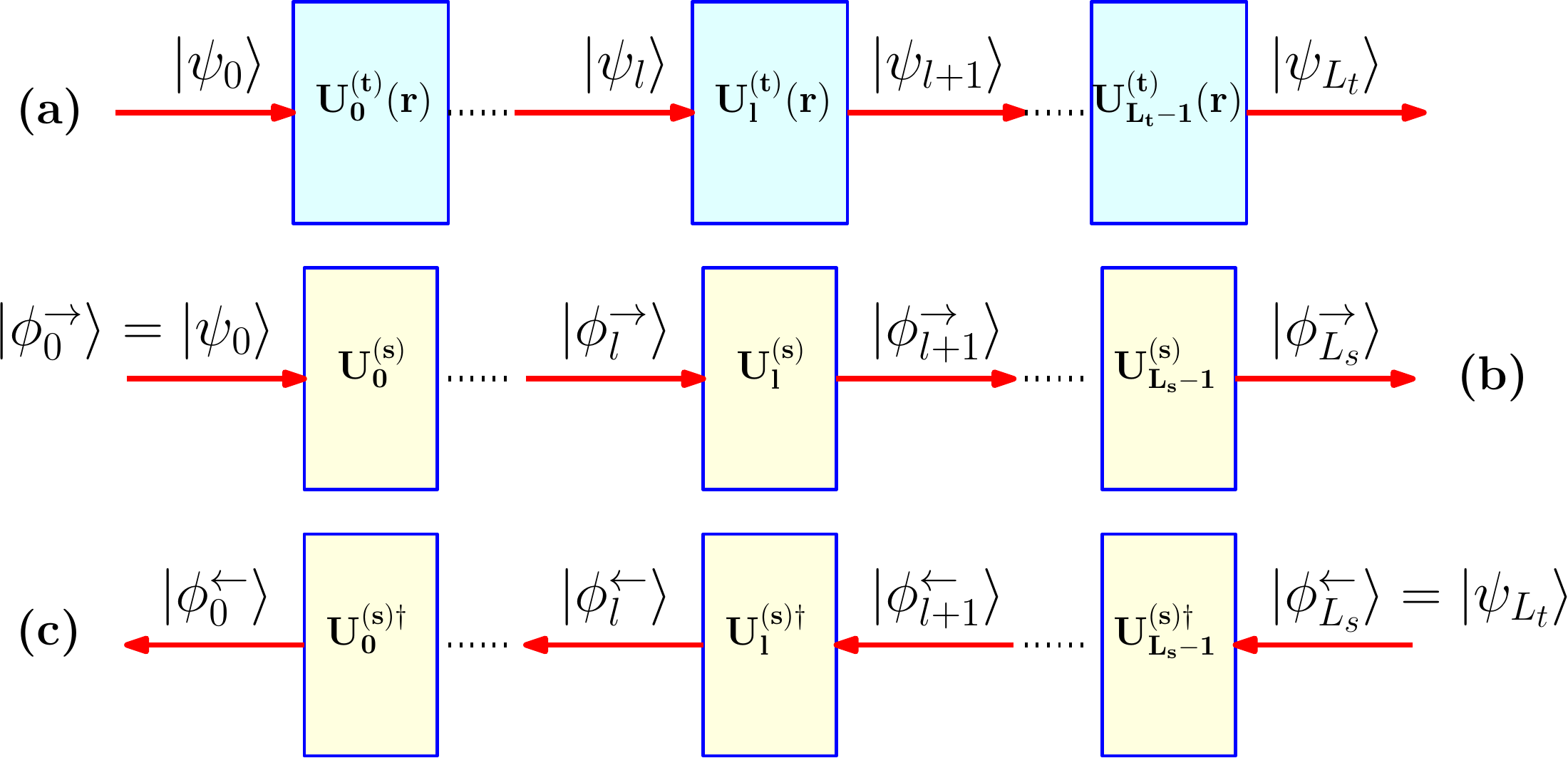} 
	\caption{Learning noise-free reinforced dynamics from the teacher. (a) The teacher generates a sequence of states $\{|\psi_l\rangle\}$ from initial state $|\psi_0\rangle$ via reinforced unitaries $U_l^{(t)}(r)$. (b) Forward stage: the student obtains $\{|\phi_l^\rightarrow\rangle\}$ from $|\phi_0^\rightarrow\rangle = |\psi_0\rangle$ by applying $U_l^{(s)}$. (c) Backward stage: the student obtains $\{|\phi_l^\leftarrow\rangle\}$ via $U_l^{(s)\dagger}$ starting from $|\phi_{L_s}^\leftarrow\rangle = |\psi_{L_t}\rangle$. The student unitaries $U_l^{(s)}$ are optimized via gradient descent to align the forward and backward dynamics.}\label{fig-M0}
\end{figure}

First, the student aims to identify the Hamiltonians $H_l^{(s)}(\boldsymbol\theta_l)$ that accurately reproduce the reinforced dynamics of the teacher in the absence of noise ($\epsilon=0$). This provides an estimate of the dynamical strategy that the teacher is working with regardless of the nature of noisy environment. It means that the learning process is independent of the noise model. Figure \ref{fig-M0} shows schematics of the learning process with the two stages of forward and backward unitary time evolutions. We employ a gradient descent algorithm which uses the information obtained by the forward and backward dynamics to infer locally the student model parameters.

More precisely, the forward time evolution of the student is obtained from $|\phi_{l+1}^{\rightarrow}\rangle =e^{-\hat{i}H_l^{(s)}}|\phi_l^{\rightarrow}\rangle$ starting with the same input $|\phi_0^{\rightarrow}\rangle=|\psi_0\rangle$ to the teacher model. In this study $|\psi_0\rangle=|\psi_i\rangle$ is the ground state of $H_i$. The backward states $|\phi_{l}^{\leftarrow}\rangle$ of the student model are given by $|\phi_{l}^{\leftarrow}\rangle =e^{\hat{i}H_l^{(s)}}|\phi_{l+1}^{\leftarrow}\rangle$.
This time the process starts with the output state in the teacher model $|\phi_{L_s}^{\leftarrow}\rangle=|\psi_{L_t}\rangle$ which is obtained by the sequence of unitary time evolutions $|\psi_{l+1}\rangle =e^{-\hat{i}H_l^{(t)}(r)}|\psi_l\rangle$.

The student aims to find the model parameters in $H_l^{(s)}(\boldsymbol\theta_l)$ which minimize the difference between the outputs of the two models. This can be done by a gradient descent (GD) algorithm where the parameters $\boldsymbol\theta_{l}$ of the student Hamiltonians are slightly changed as follows 
\begin{align}
\Delta \boldsymbol\theta_{l} =-\eta \frac{\partial}{\partial \boldsymbol\theta_{l}}e_l,
\end{align}
with the learning rate $\eta>0$. The local error functions $e_l$ are given by  
\begin{align}
e_l &=\frac{1}{2}\langle \Delta\phi_{l+1}|\Delta\phi_{l+1}\rangle,\\
|\Delta \phi_l\rangle &=|\phi_l^{\leftarrow}\rangle-|\phi_l^{\rightarrow}\rangle,
\end{align}
to measure the consistency of the forward and backward dynamics at layer $l$.  
In each iteration of the GD algorithm we update all parameters in the student model once. The final error of the learning process then reads $e =\frac{1}{2}\langle \Delta\phi_{L_s}|\Delta\phi_{L_s}\rangle$. 

The student then uses the inferred Hamiltonians for its evolution now in presence of a noisy environment.   
Given an input state $\rho_0$ to the student, the sequence of states are given by $\rho_{l+1}=\mathcal{E}_l(U_{l}^{(s)}\rho_l U_{l}^{(s)\dagger})$ up to the output state $\rho_{L_s}$.  Starting with the ground state of $H_i$, $\rho_0=|\psi_i \rangle\langle \psi_i|$, the success probability of the student at the end is $P_{success}^{(s)}=\mathrm{Tr}(\rho_{L_s}|\psi_f \rangle\langle \psi_f|)$. The channel $\mathcal{E}_l$ affects the student and teacher states in the same way as defined in the previous section.

In practice, we limit the structure and strength of interactions in the student model to have a local and physically economical Hamiltonian. The number of parameters in such models is proportional to size of system and number of layers, $LN$. The gradients of $e_l$ with respect to the $O(N)$ parameters in layer $l$ can be computed numerically using on the order of $N$ unitary evolutions.  The effects of changes in the parameters of layer $l$ are then propagated through the system with approximately $L$ unitary evolutions. Unitary evolutions generated by local and sparse Hamiltonians can be approximated using $O(N)$ local unitary gates. Therefore, for a model with $L$ layers, approximately $L[N(N+LN)]$ local unitary operations are needed.  Requiring exact unitary evolutions, of course, increases the time complexity exponentially, from $O(L^2N^2)$ to $O(L^2Ne^N)$.

\section{Reinforced dynamics of a single qubit}\label{S3}
A quantum search problem can be naturally framed as an annealing process that evolves the system from the ground state of an initial Hamiltonian, $H_i=\mathbb{I}-|\psi_i \rangle\langle \psi_i|$ to the ground state of a final Hamiltonian, $H_f=\mathbb{I}-|\psi_f \rangle\langle \psi_f|$. Here, the initial state $|\psi_i \rangle$ is prepared as a superposition of the solution subspace $|\psi_f \rangle$ and its orthogonal complement $|\psi_f ^{\perp}\rangle$. That is $|\psi_i \rangle=\sqrt{P_0}|\psi_f \rangle+\sqrt{1-P_0}|\psi_f^{\perp} \rangle$, where $P_0$ denotes the initial probability of measuring a solution. In our analysis, we fix $P_0=2^{-10}$ to reflect a typical setting where the solution is initially rare. In a standard annealing process with $H_l^{(t)}(0)=(1-t_l)H_i+t_lH_f$, the system dynamics reduces to that of an effective two-level system (qubit). In this section, we study the reinforced dynamics of such a two dimensional ($d=2$) system under the influence of a noisy environment. That is we assume that both reinforcement and noise act only at the level of the two subspaces in the quantum search problem. For the annealing schedule we use the optimal form given in Eq. \ref{grover}.

The student Hamiltonians are parameterized in terms of Pauli matrices as $H_l^{(s)}(\boldsymbol\theta_l)=\sum_{\mu}\theta_{l,\mu}\sigma_{\mu}$ where
$\sigma_{\mu=x,y,z}$ denote the Pauli operators and $\theta_{l,\mu}$ are the corresponding coefficients. The computational basis states $|\sigma=\pm\rangle$ - eigenstates of $\sigma_z$ with eigenvalues $\pm 1$ - are used to represent the solution subspace $|\psi_f \rangle$ and its orthogonal complement $|\psi_f^{\perp} \rangle$, respectively. The magnitudes of the Hamiltonian coefficients are constrained to $|\theta_{l,\mu}|<1$ to prevent costly terms in the Hamiltonian that could hinder experimental feasibility or numerical stability. The initial parameters $\theta_{l,\mu}$ are sampled randomly and uniformly from the interval $(-1,+1)$, providing an unbiased starting point for the optimization process. For training, we employ a learning rate of $\eta=1$, chosen to balance rapid convergence with stability during the GD updates.

\begin{figure}
	\includegraphics[width=16cm]{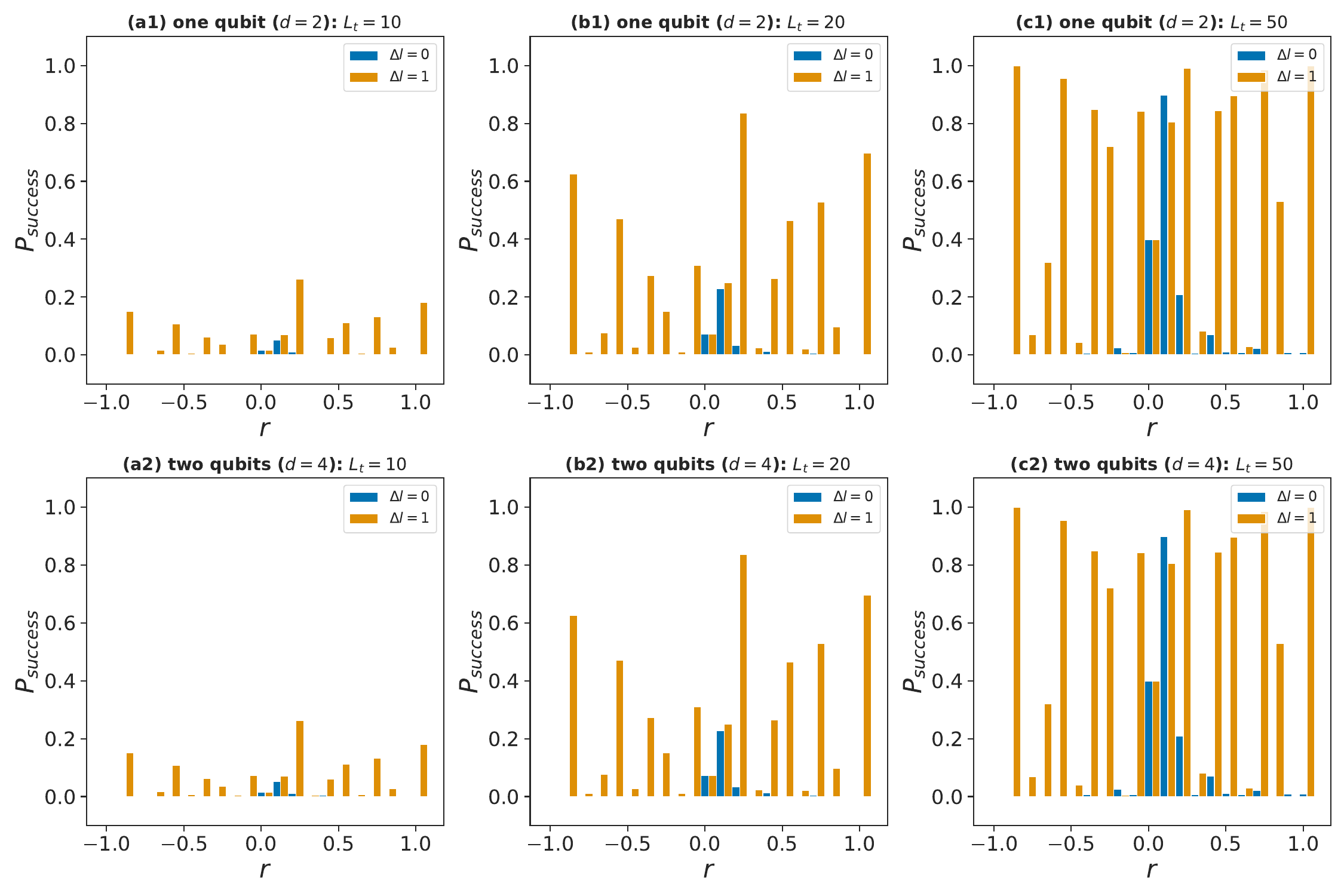} 
	\caption{Success probability of the student model under noise-free conditions. ((a1), (b1), (c1)) For a single qubit ($d=2$). ((a2), (b2), (c2)) For two qubits ($d=4$). The number of student layers is $L_s=5$. The results are obtained after $100$ GD iterations with learning rates $\eta=1$ and $\eta=0.02$ for $d=2$ and $d=4$, respectively.}\label{fig-M1}
\end{figure}

\begin{figure}
	\includegraphics[width=16cm]{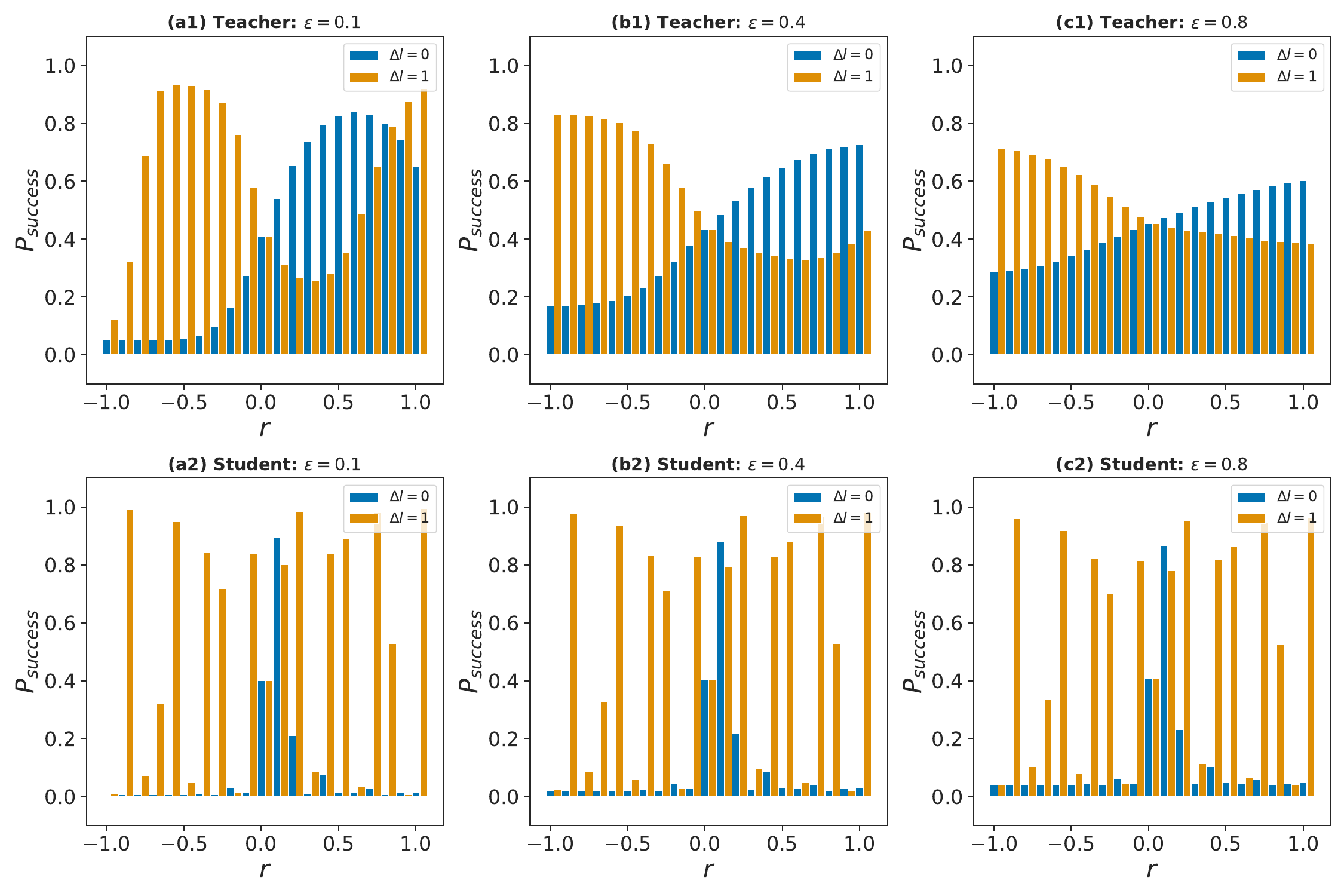} 
	\caption{One qubit: Success probability under depolarizing noise. Here $L_t=50$ and $L_s=5$. ((a1), (b1), (c1)) For the teacher model. ((a2), (b2), (c2)) For the student model. The student results are obtained after $100$ GD iterations with learning rate $\eta=1$.}\label{fig-M2}
\end{figure}

Figure \ref{fig-M1} (upper panels) displays success probability of the student after learning the reinforced dynamics of the teacher in the absence of noise. We see how $P_{success}^{(s)}$ changes with the strength of reinforcement $r$ and $\Delta l$ for different numbers of teacher layers $L_t=10, 20, 50$. The number of layers in the student model is fixed to $L_s=5$; we try to find an effective representation of the reinforced dynamics with a smaller number of ordinary (not reinforced) evolution steps. This allows us to reduce also the total computation time and therefore the effects of noisy interactions with the environment. The learning algorithm gives an error of order $< 10^{-6}$ after $100$ GD iterations. Details of the GD algorithm are given in Appendix \ref{app-A}. Figure \ref{fig-M1} shows that the highest success probability is obtained for nonzero reinforcements. This means that even in the absence of noise reinforcement enhances success probability of the search algorithm, as expected from previous studies \cite{QR-pra-2022}. Moreover, larger values of $\Delta l$ result in better performances specially for small number of layers $L_t$. For large $L_t$, the success probability is already high when $\Delta l=0$ which leaves little room for improvement.

\begin{figure}
	\includegraphics[width=12cm]{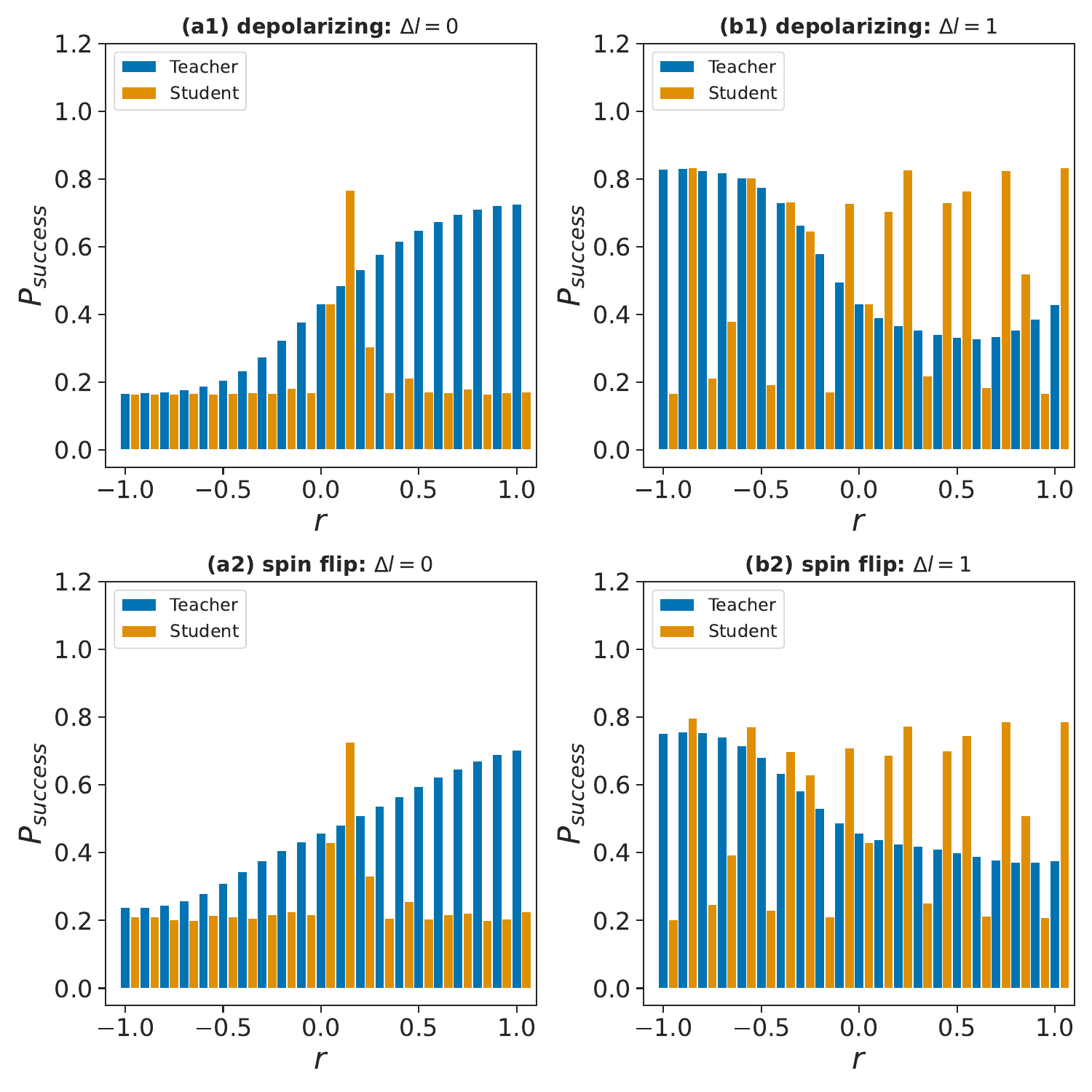} 
	\caption{One qubit: Success probability under depolarizing and bit-flip noise. Here $L_t=L_s=50$, $\epsilon=0.4$. ((a1), (b1)) For depolarizing noise. ((a2), (b2)) For bit-flip noise. The student results are obtained after $100$ GD iterations with learning rate $\eta=1$.}\label{fig-M3}
\end{figure}

To analyze the performance of the student and teacher models under noisy conditions, we consider the depolarizing noise from Pauli channel as our primary noise model, 
\begin{align}
\mathcal{E}_l(\rho) =(1-\epsilon_l)\rho+\frac{\epsilon_l}{d}\mathbb{I},
\end{align}
where $\mathbb{I}$ is the identity operator. The depolarizing channel provides a well-understood framework for studying decoherence effects in quantum systems. Figure \ref{fig-M2} displays $P_{success}$ as a function of reinforcement strength $r$ for the two models under varying noise strengths $\epsilon$. The analysis uses $L_t=50$, $L_s=5$, and considers $\Delta l = 0,1$ cases. Additional results for $L_t=10,20$ are provided in Appendix \ref{app-A}. We emphasize that by increasing the number of layers in the student model we increase the model susceptibility to noise. Specially when $L_s=L_t$ we expect to observe the sole impact of reinforcement on the quantum dynamics. This behavior is shown in Fig. \ref{fig-M3} for $L_t=L_s=50$ and $\epsilon=0.4$. Besides the depolarizing noise, the figure also shows the success probabilities for the bit-flip noise
\begin{align}
\mathcal{E}_l(\rho) =(1-\epsilon_l)\rho+\epsilon_l\sigma_x\rho\sigma_x.
\end{align}
As the figure shows, the models exhibit similar performances under both depolarizing and bit-flip noise, suggesting robustness to variations in noise type. Note that here we have no randomness in the unitary and Kraus operators. Thus there is no statistical error in the reported data.

\section{Reinforced dynamics of two qubits}\label{S4}
In this section, we study a four-state ($d=4$) system consisting of two qubits. This allows us to have a more accurate picture of reinforcement and noise effects in the quantum search problem. 
Here the quantum states are represented in the computational basis of the two qubits $|\sigma\sigma'\rangle$ with 
\begin{align}
|\psi_i\rangle=\sqrt{P_0}|++\rangle+\sqrt{\frac{P_0}{3}}(|+-\rangle+|-+\rangle+|--\rangle),
\end{align}
$|\psi_f\rangle=|++\rangle$, and $P_0=2^{-10}$. The primed variables refer to the second qubit.
The Hamiltonians in each layer of the student model are
\begin{align}
H_l^{(s)}=\sum_{\mu,\mu'}\theta_{l,\mu\mu'}\sigma_{\mu} \sigma_{\mu'}',
\end{align}
where $\mu=0,x,y,z$ and $\sigma_0$ is the identity matrix. As before, the magnitude of parameters are restricted to $|\theta_{l,\mu\mu'}|<1$. The initial parameters are chosen randomly and uniformly in $(-10^{-6},+10^{-6})$. The annealing schedule is the optimal one given in Eq. \ref{grover}, and the learning rate $\eta=0.02$.

\begin{figure}
	\includegraphics[width=16cm]{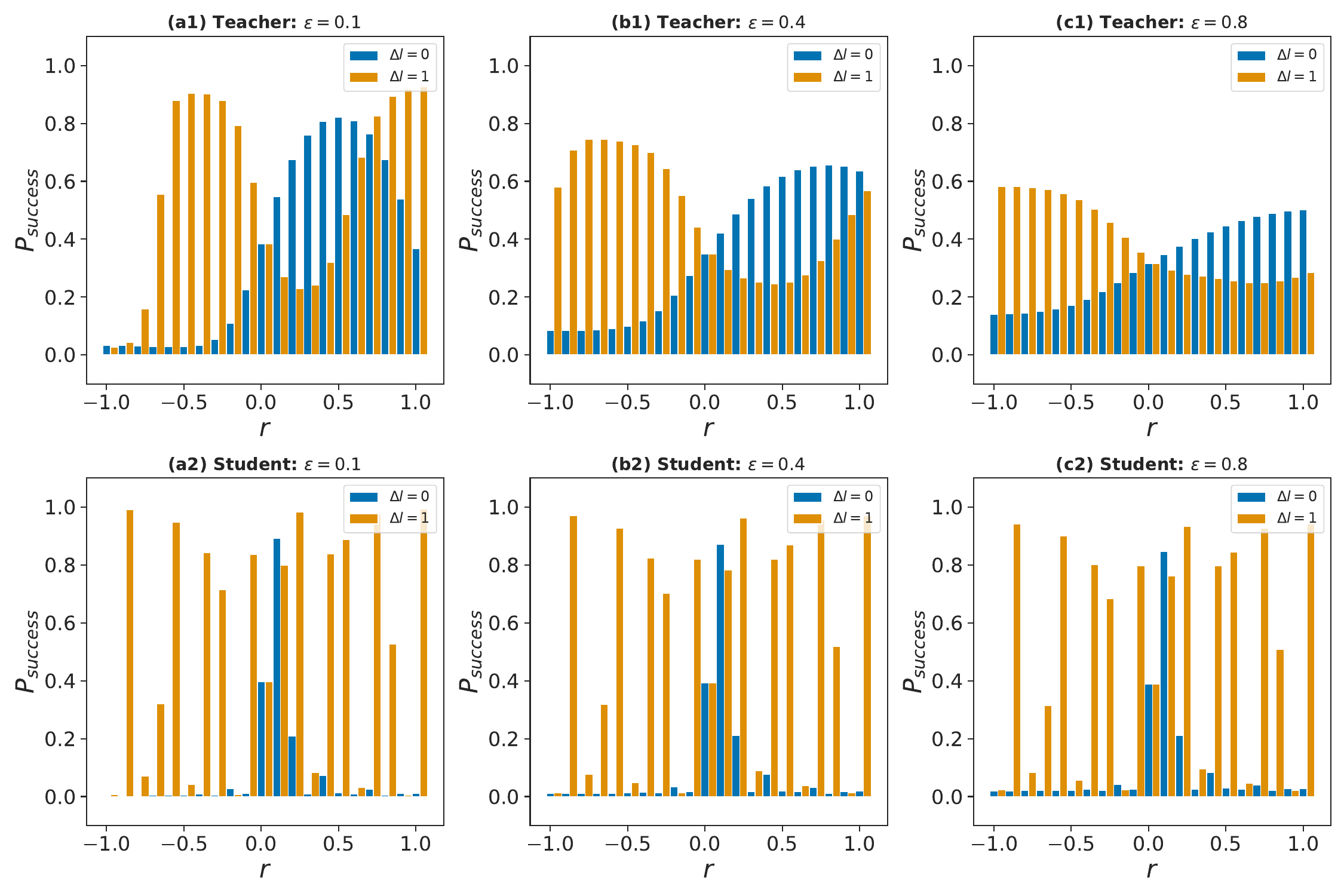} 
	\caption{Two qubits: Success probability under depolarizing noise. Here $L_t=50$ and $L_s=5$. ((a1), (b1), (c1)) For the teacher model. ((a2), (b2), (c2)) For the student model. The student results are obtained after $100$ GD iterations with learning rate $\eta=0.02$.}\label{fig-M4}
\end{figure}

The lower panels of Fig. \ref{fig-M1} show success probability of the student model after learning the reinforced dynamics of the teacher in the noiseless case ($\epsilon=0$). Here we examine how $P_{success}$ varies with the reinforcement strength $r$ and $\Delta l$ for $L_t = 10, 20, 50$, while keeping $L_s = 5$. The learning algorithm converges within a few tens of gradient descent iterations, achieving an error $< 10^{-17}$. Computational details are provided in Appendix \ref{app-B}. Note that the optimal success probability is obtained for a nonzero reinforcement and its value is nearly two times larger than the probability we obtain without reinforcement for $r=0$.    

As before, we consider both depolarizing and bit-flip noise models to simulate the noisy evolution of the system. Figure \ref{fig-M4} illustrates the dependence of $P_{success}$ on $r$ for selected noise strengths $\epsilon$, using depolarizing noise with $L_t = 50$ and $L_s = 5$. Additional results for other $L_t$ values and the bit-flip noise model are included in Appendix \ref{app-B}. As Fig. \ref{fig-M4} shows, larger magnitudes of reinforcement are needed to maximize the success probability as the noise strength increases in the teacher model. The student behavior does not change too much with the noise level because $L_s=5$ is very small compared to $L_t=50$. Notably, the qualitative behaviors remain consistent between single- and two-qubit systems, showing minimal sensitivity to system size. We expect to observe similar performances for a larger number of qubits; previous studies of reinforcement algorithms in other optimization problems show that the technique can be helpful also in the limit of large system sizes \cite{baldassi-pnas-2007,QR-pra-2022}.

\section{Conclusion}\label{S5}
We observed that quantum reinforcement enhances noise tolerance in a quantum annealing simulation. This was achieved by reinforcing the system dynamics by the quantum state of the system. A learning algorithm was employed to minimize the physical cost of reinforcement and reduce the total evolution time in a noisy environment. Future work could explore the performance of both the original (teacher) and approximate (student) algorithms in other optimization problems—particularly for larger system sizes and experimental implementations. Additionally, one could improve the student model by learning the noisy dynamics of the teacher model, enabling better approximations tailored to specific noise profiles. Finally, measurement effects and state estimation errors must be more accurately accounted for in the teacher reinforced dynamics.

Two main limitations can be identified in the method proposed in this study. First, at each layer of the teacher dynamics, we need to know the system’s quantum state from the preceding layer. In principle, collective weak measurements could be employed to obtain this information, but such procedures are computationally and experimentally costly. Second, the learning algorithm developed for the student model to circumvent the above issue requires $L^2N$ unitary evolutions, whose exact treatment scales exponentially with $N$. Although approximate treatments of these unitary evolutions can reduce the time complexity to 
$O(N)$ for local and sparse Hamiltonians, they may simultaneously degrade the learning performance and reduce the success probability of the student model. Further investigations are therefore required to develop strategies that can mitigate these challenges.

\acknowledgments
I acknowledge useful discussions with Marjan Homayouni-Sangari who did preliminary investigations of the learning algorithm during her Master thesis. I would like to thank Mohammad Hossein Zarei for helpful discussions.  
This work was performed using the ALICE compute resources provided by Leiden University.

\appendix
\counterwithin{figure}{section}

\section{Learning reinforced dynamics of a single qubit}\label{app-A}
For the teacher we consider $L_t$ layers of noisy time evolutions of a single qubit with quantum states $\rho_l$ (Fig. \ref{fig-M000}),
\begin{align}
\rho_{l+1} &=\mathcal{E}_l(U_l^{(t)}(r)\rho_lU_l^{(t)\dagger}(r)),\\
U_l^{(t)}(r) &=e^{-\hat{i}H_l^{(t)}(r)}.
\end{align}
We consider two kinds of noise models from Pauli channel. The depolarizing noise,
\begin{align}
\mathcal{E}_l(\rho) =(1-\epsilon_l)\rho+\frac{\epsilon_l}{2}\mathbb{I},
\end{align}
and the bit-flip noise,
\begin{align}
\mathcal{E}_l(\rho) =(1-\epsilon_l)\rho+\epsilon_l \sigma_x\rho\sigma_x.
\end{align}
$\sigma_{x,y,z}$ are Pauli matrices and $\epsilon_l=\epsilon/L_t$ is the noise strength. We work with the eigenstates $|\sigma=\pm\rangle$ of $\sigma_z$ as the computational basis. 

The teacher performs a reinforced quantum annealing to go from the ground state of initial Hamiltonian $H_i$ to that of final Hamiltonian $H_f$,
\begin{align}
H_l^{(t)}(r) &=(1-t_l)H_i+t_lH_f-r_l\ln\rho_{l+\Delta l}(0),\\ 
H_i &=1-|\psi_i \rangle\langle \psi_i|,\\
H_f &=1-|\psi_f \rangle\langle \psi_f|,\\
|\psi_i \rangle &=\sqrt{P_0}|+ \rangle+\sqrt{1-P_0}|- \rangle,\\
|\psi_f \rangle &=|+ \rangle.
\end{align}
$P_0$ controls the initial overlap with the target state $|+\rangle$. In this study we set $P_0=2^{-10}$.
The teacher dynamics is reinforced by the state $\rho_{l+\Delta l}(0)$ which is expected to be observed in the absence of reinforcement and noise,
\begin{align}
\rho_{l+1}(0) &=U_l^{(t)}(0)\rho_{l}U_l^{(t)\dagger}(0),\\
\rho_{l+2}(0) &=U_l^{(t)}(0)\rho_{l+1}(0)U_l^{(t)\dagger}(0),\dots
\end{align}
starting with the current state $\rho_{l}$.
We assume that the reinforcement parameter $r_l=r$ does not change with time $t_l\in (0,1)$. Moreover, we consider the optimal annealing schedule $t_l^*$ of the Grover algorithm,
\begin{align}
t_l^* &=\frac{1}{2}\left[1-\sqrt{\frac{P_0}{1-P_0}}\tan\left((1-2\frac{l}{L_t-1})\alpha\right)\right],\\
\alpha &=\arctan\left(\sqrt{\frac{1-P_0}{P_0}}\right).
\end{align}
The success probability of the teacher is measured by $P_{success}^{(t)}=\mathrm{Tr}(\rho_{L_t}|+ \rangle\langle +|)$.

The student also performs $L_s$ layers of noisy time evolutions of a single qubit with quantum states $\rho_l$,
\begin{align}
\rho_{l+1} &=\mathcal{E}_l(U_l^{(s)}\rho_lU_l^{(s)\dagger}),\\
U_l^{(s)} &=e^{-\hat{i}H_l^{(s)}}.
\end{align}
The noise models are the same as the noise experienced by the teacher model.
Success probability of the student is $P_{success}^{(s)}=\mathrm{Tr}(\rho_{L_s}|+ \rangle\langle +|)$.
The Hamiltonians $H_l^{(s)}$ are obtained by learning the reinforced dynamics of the teacher in the absence of noise as follows.

As a single qubit, the student Hamiltonian in each layer in general is
\begin{align}
H=\mathbf{h}.\boldsymbol\sigma=h_{x}\sigma_x+h_{y}\sigma_y+h_{z}\sigma_z,
\end{align}
with parameters $\theta_{x,y,z}=h_{x,y,z}$. 
The time evolution operator then is given by  
\begin{align}
U=e^{-\hat{i}H}=\left(\begin{matrix}
\cos(h)-\hat{i}\frac{h_z}{h}\sin(h) & \frac{-h_y-\hat{i}h_x}{h}\sin(h) \\
\frac{h_y-\hat{i}h_x}{h}\sin(h) & \cos(h)+\hat{i}\frac{h_z}{h}\sin(h)
\end{matrix}\right),
\end{align}
where $h^2=h_{x}^2+h_{y}^2+h_{z}^2$. 

Suppose the teacher obtains the output state $|\psi_{L_t}\rangle$ starting with the ground state of $H_i$, that is $|\psi_0\rangle=|\psi_i\rangle$ (Fig. \ref{fig-M0}). The forward dynamics of the student gives $|\phi_{l+1}^{\rightarrow}\rangle=e^{-\hat{i}H_l^{(s)}}|\phi_{l}^{\rightarrow}\rangle$ starting with $|\phi_{0}^{\rightarrow}\rangle=|\psi_0\rangle$.
The backward dynamics of the student gives $|\phi_{l}^{\leftarrow}\rangle=e^{+\hat{i}H_l^{(s)}}|\phi_{l+1}^{\leftarrow}\rangle$ starting with $|\phi_{L_s}^{\leftarrow}\rangle=|\psi_{L_t}\rangle$.

Given the forward and backward states,
\begin{align}
|\phi_l^{\rightarrow}\rangle=\alpha_+|+\rangle+\alpha_-|-\rangle,\\
|\phi_{l+1}^{\leftarrow}\rangle=\beta_+|+\rangle+\beta_-|-\rangle,
\end{align}
the error function at layer $l$ reads
\begin{align}
e_l &=\frac{1}{2}\langle \Delta\phi_{l+1}|\Delta\phi_{l+1}\rangle,\\
|\Delta \phi_{l+1}\rangle &=|\phi_{l+1}^{\leftarrow}\rangle-|\phi_{l+1}^{\rightarrow}\rangle=|\phi_{l+1}^{\leftarrow}\rangle-e^{-\hat{i}H_l^{(s)}}|\phi_l^{\rightarrow}\rangle,\\
e_l &=1-\mathrm{Re}[f_l],\\
f_l &=\langle \phi_{l+1}^{\leftarrow}|e^{-\hat{i}H_l^{(s)}}|\phi_{l}^{\rightarrow}\rangle.
\end{align}

In terms of the model parameters
\begin{align}
f_l &=A\cos(h)-\hat{i}\mathbf{B}.\frac{\mathbf{h}}{h}\sin(h),\\
A &= \alpha_+\beta_+^* + \alpha_-\beta_-^*,\\
B_x &= \alpha_+\beta_-^* + \alpha_-\beta_+^*,\\
B_y &=\hat{i}[\alpha_+\beta_-^* - \alpha_-\beta_+^*],\\
B_z &=\alpha_+\beta_+^* - \alpha_-\beta_-^*.
\end{align}

\begin{figure}
	\includegraphics[width=16cm]{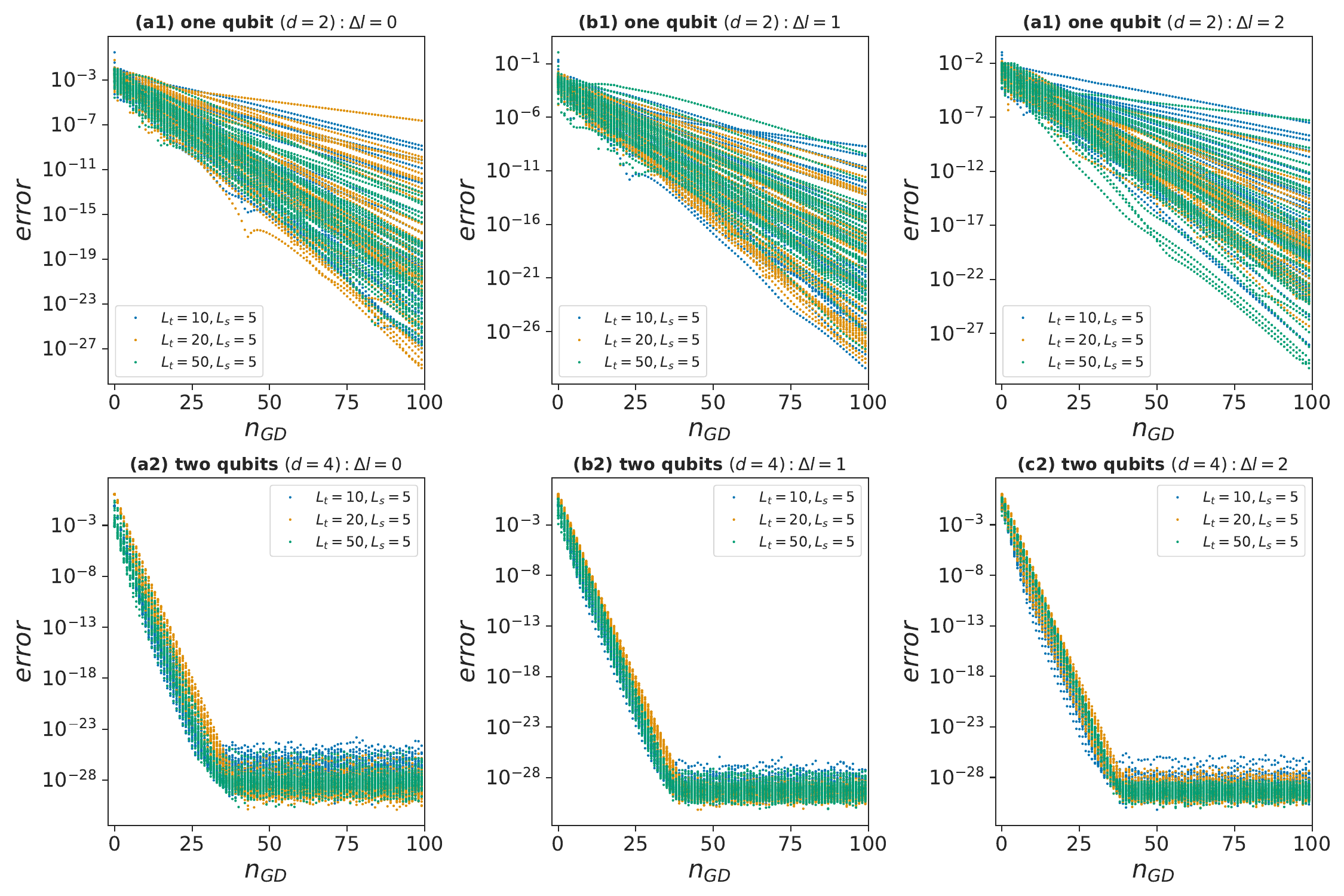} 
	\caption{Student's learning error during training. ((a1), (b1), (c1)) For a single qubit ($d=2$) with learning rate $\eta=1$. ((a2), (b2), (c2)) For two qubits ($d=4$) with learning rate $\eta=0.02$. The numbers of teacher and student layers are $L_t=10, 20, 50$ and $L_s=5$. The data are for $r\in (-1,+1)$ in the teacher model.}\label{fig-A0}
\end{figure}

\begin{figure}
	\includegraphics[width=16cm]{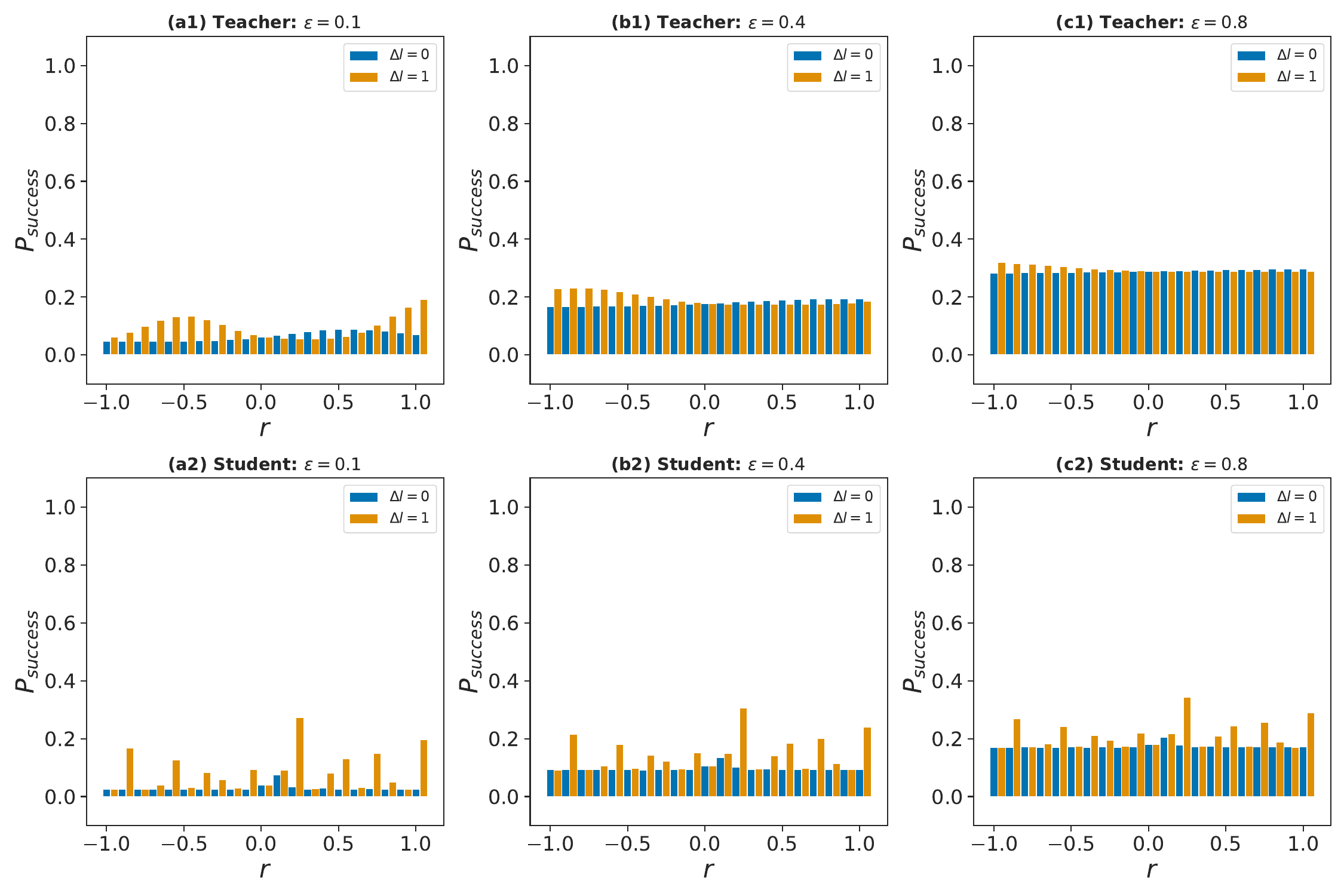} 
	\caption{One qubit: Success probability under depolarizing noise. Here $L_t=10$ and $L_s=5$. ((a1), (b1), (c1)) For the teacher model. ((a2), (b2), (c2)) For the student model. The student results are obtained after $100$ GD iterations with learning rates $\eta=1$.}\label{fig-A1}
\end{figure}

\begin{figure}
	\includegraphics[width=16cm]{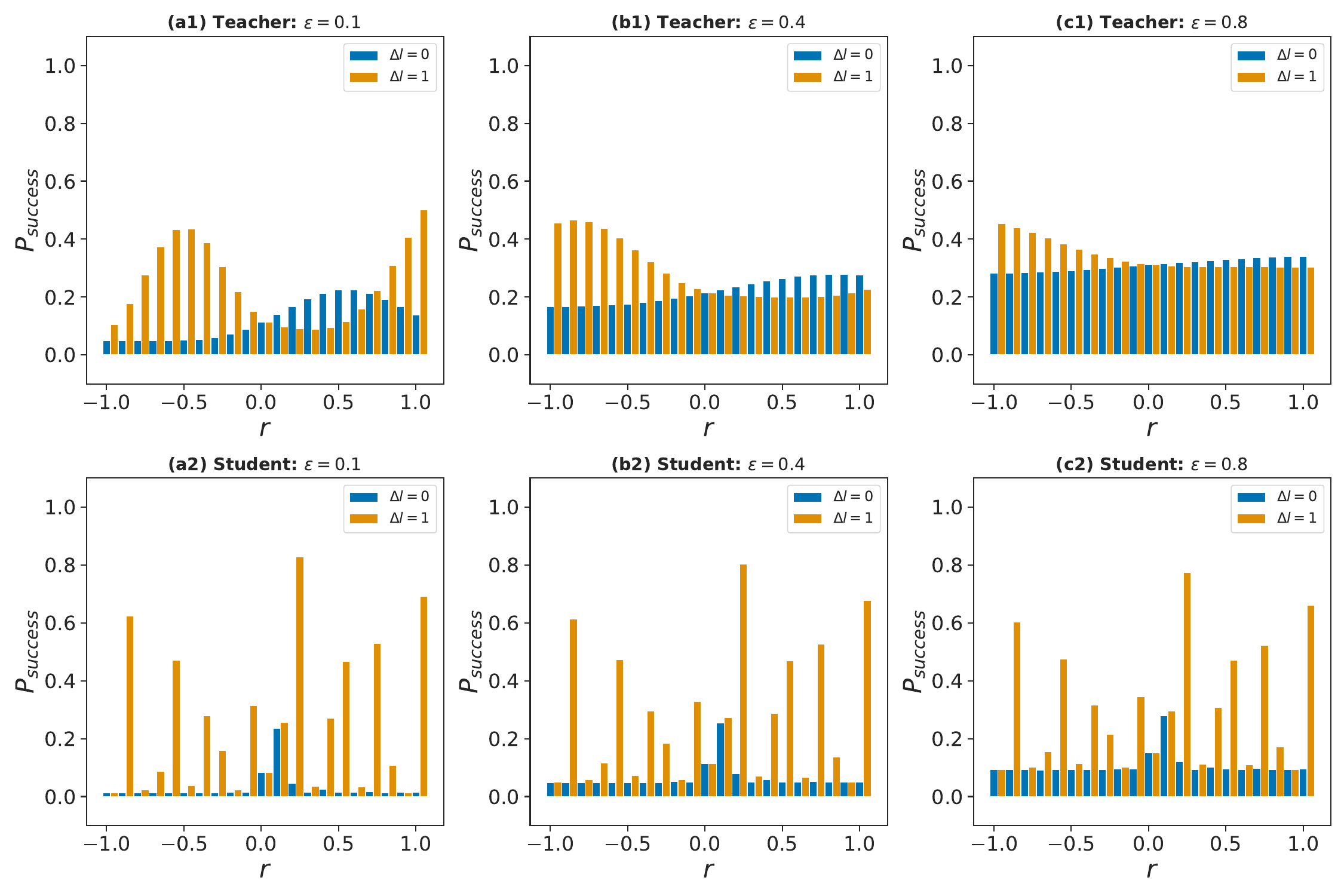} 
	\caption{One qubit: Success probability under depolarizing noise. Here $L_t=20$ and $L_s=5$. ((a1), (b1), (c1)) For the teacher model. ((a2), (b2), (c2)) For the student model. The student results are obtained after $100$ GD iterations with learning rates $\eta=1$.}\label{fig-A2}
\end{figure}

Derivatives of the error function $e_l$ then are 
\begin{align}
\frac{\partial e_l}{\partial h_x} &=\frac{h_x}{\sqrt{h}}\left([\mathrm{Re}(A)+\frac{\mathbf{h}}{h^2}.\mathrm{Im}(\mathbf{B})]\sin(h)-\frac{\mathbf{h}}{h}.\mathrm{Im}(\mathbf{B})\cos(h)\right)-\frac{\mathrm{Im}(B_x)}{h}\sin(h),\\
\frac{\partial e_l}{\partial h_y} &=\frac{h_y}{\sqrt{h}}\left([\mathrm{Re}(A)+\frac{\mathbf{h}}{h^2}.\mathrm{Im}(\mathbf{B})]\sin(h)-\frac{\mathbf{h}}{h}.\mathrm{Im}(\mathbf{B})\cos(h)\right)-\frac{\mathrm{Im}(B_y)}{h}\sin(h),\\
\frac{\partial e_l}{\partial h_z} &=\frac{h_z}{\sqrt{h}}\left([\mathrm{Re}(A)+\frac{\mathbf{h}}{h^2}.\mathrm{Im}(\mathbf{B})]\sin(h)-\frac{\mathbf{h}}{h}.\mathrm{Im}(\mathbf{B})\cos(h)\right)-\frac{\mathrm{Im}(B_z)}{h}\sin(h).
\end{align}
The model parameters are modified by a gradient descent (GD) algorithm
\begin{align}
\Delta \boldsymbol\theta_{l} =-\eta \frac{\partial}{\partial \boldsymbol\theta_{l}}e_l,
\end{align}
with the learning rate $\eta$. In each iteration of the GD algorithm all parameters are updated once.

The main quantities are the error function of the learning process
\begin{align}
e &=\frac{1}{2}\langle \Delta\phi_{L_s}|\Delta\phi_{L_s}\rangle,\\
|\Delta \phi_{L_s}\rangle &=|\phi_{L_s}^{\leftarrow}\rangle-|\phi_{L_s}^{\rightarrow}\rangle,
\end{align} 
and the success probabilities of the teacher and student models. Figure \ref{fig-A0} (upper panels) shows the error values with parameters $h_{x,y,z}\in (-1,+1)$. The initial values also lie in $(-1,+1)$, with a learning rate $\eta=1$. 
Figures \ref{fig-A1} and \ref{fig-A2} display the success probabilities for different values of $r, \Delta l, \epsilon$ when $L_t=10, 20$ and $L_s=5$.

\section{Learning reinforced dynamics of two qubits}\label{app-B}
The teacher model represents dynamics of two qubits with $L_t$ layers of noisy time evolutions of quantum states $\rho_l$ (Fig. \ref{fig-M000}),
\begin{align}
\rho_{l+1} &=\mathcal{E}_l(U_l^{(t)}(r)\rho_lU_l^{(t)\dagger}(r)),\\
U_l^{(t)}(r) &=e^{-\hat{i}H_l^{(t)}(r)}.
\end{align}
Here the depolarizing noise is
\begin{align}
\mathcal{E}_l(\rho) =(1-\epsilon_l)\rho+\frac{\epsilon_l}{4}\mathbb{I},
\end{align}
and the bit-flip noise is
\begin{align}
\mathcal{E}_l(\rho) =(1-\epsilon_l)\rho+\frac{\epsilon_l}{3}(\sigma_x\rho\sigma_x+\sigma_x'\rho\sigma_x'+\sigma_x\sigma_x'\rho\sigma_x\sigma_x').
\end{align}

The quantum states are represented in the computational basis of the two qubits $|\sigma\sigma'\rangle$.
As before, we consider a reinforced quantum annealing process with the follwing time-dependent Hamiltonian 
\begin{align}
H_l^{(t)}(r) &=(1-t_l)H_i+t_lH_f-r_l\ln\rho_{l+\Delta l}(0),\\ 
H_i &=1-|\psi_i \rangle\langle \psi_i|,\\
H_f &=1-|\psi_f \rangle\langle \psi_f|,\\
|\psi_i \rangle &=\sqrt{P_0}|+ +\rangle+\sqrt{\frac{1-P_0}{3}}(|+-\rangle+|-+\rangle+|--\rangle),\\
|\psi_f \rangle &=|+ +\rangle.
\end{align}
$P_0$ controls the initial overlap with the target state $|++\rangle$. In this study we set $P_0=2^{-10}$.
The teacher dynamics is reinforced by the state $\rho_{l+\Delta l}(0)$ which is obtained in the absence of reinforcement and noise,
\begin{align}
\rho_{l+1}(0) &=U_l^{(t)}(0)\rho_{l}U_l^{(t)\dagger}(0),\\
\rho_{l+2}(0) &=U_l^{(t)}(0)\rho_{l+1}(0)U_l^{(t)\dagger}(0),\dots
\end{align}
starting with the current state $\rho_{l}$ at layer $l$.
The reinforcement parameter $r_l=r$ does not change with time $t_l\in (0,1)$. Moreover, we work with the optimal annealing schedule $t_l^*$ of the Grover algorithm,
\begin{align}
t_l^* &=\frac{1}{2}\left[1-\sqrt{\frac{P_0}{1-P_0}}\tan\left((1-2\frac{l}{L_t-1})\alpha\right)\right],\\
\alpha &=\arctan\left(\sqrt{\frac{1-P_0}{P_0}}\right).
\end{align}
The success probability of the teacher is measured by $P_{success}^{(t)}=\mathrm{Tr}(\rho_{L_t}|+ +\rangle\langle + +|)$.

The student also performs $L_s$ layers of noisy time evolutions on two qubits of quantum state $\rho_l$,
\begin{align}
\rho_{l+1} &=\mathcal{E}_l(U_l^{(s)}\rho_lU_l^{(s)\dagger}),\\
U_l^{(s)} &=e^{-\hat{i}H_l^{(s)}},
\end{align}
with success probability $P_{success}^{(s)}=\mathrm{Tr}(\rho_{L_s}|+ +\rangle\langle + +|)$.
The Hamiltonians $H_l^{(s)}$ are obtained by learning the reinforced dynamics of the teacher in absence of any noise as follows.

The Hamiltonians in each layer of the student model are in general
\begin{align}
H_l^{(s)}=\sum_{\mu,\mu'}\theta_{l,\mu\mu'}\sigma_{\mu} \sigma_{\mu'}',
\end{align}
where $\mu=0,x,y,z$ and $\sigma_0$ is the identity matrix. 

Suppose the teacher obtains the output state $|\psi_{L_t}\rangle$ starting with the ground state of $H_i$, that is $|\psi_0\rangle=|\psi_i\rangle$ (Fig. \ref{fig-M0}). The forward dynamics of the student gives $|\phi_{l+1}^{\rightarrow}\rangle=e^{-\hat{i}H_l^{(s)}}|\phi_{l}^{\rightarrow}\rangle$ starting with $|\phi_{0}^{\rightarrow}\rangle=|\psi_0\rangle$.
The backward dynamics of the student gives $|\phi_{l}^{\leftarrow}\rangle=e^{+\hat{i}H_l^{(s)}}|\phi_{l+1}^{\leftarrow}\rangle$ starting with $|\phi_{L_s}^{\leftarrow}\rangle=|\psi_{L_t}\rangle$.

Given the forward and backward states, the error function at layer $l$ reads
\begin{align}
e_l &=\frac{1}{2}\langle \Delta\phi_{l+1}|\Delta\phi_{l+1}\rangle,\\
|\Delta \phi_{l+1}\rangle &=|\phi_{l+1}^{\leftarrow}\rangle-|\phi_{l+1}^{\rightarrow}\rangle=|\phi_{l+1}^{\leftarrow}\rangle-e^{-\hat{i}H_l^{(s)}}|\phi_l^{\rightarrow}\rangle.
\end{align}
The model parameters are modified by a gradient descent (GD) algorithm
\begin{align}
\Delta \boldsymbol\theta_{l} =-\eta \frac{\partial}{\partial \boldsymbol\theta_{l}}e_l,
\end{align}
with the learning rate $\eta$. 

Recall that
\begin{align}
e_l &=1-\mathrm{Re}[f_l],\\
f_l &=\langle \phi_{l+1}^{\leftarrow}|U_l^{(s)}|\phi_{l}^{\rightarrow}\rangle,\\
\frac{\partial f_l}{\partial \boldsymbol\theta_{l}} &=\langle \phi_{l+1}^{\leftarrow}|\frac{\partial}{\partial \boldsymbol\theta_{l}}U_l^{(s)}|\phi_{l}^{\rightarrow}\rangle,
\end{align}
Let us rewrite
\begin{align}
U_l^{(s)} =\sum_{\mu,\mu'}\mathrm{Tr}(\sigma_{\mu} \sigma_{\mu'}' U_l^{(s)})\sigma_{\mu} \sigma_{\mu'}'.
\end{align}
Then
\begin{align}
\frac{\partial}{\partial \boldsymbol\theta_{l}}U_l^{(s)} &=-\hat{i}\sum_{\mu,\mu'}\mathrm{Tr}(\sigma_{\mu} \sigma_{\mu'}' \frac{\partial H_l^{(s)}}{\partial \boldsymbol\theta_{l}}U_l^{(s)})\sigma_{\mu} \sigma_{\mu'}',\\
\frac{\partial H_l^{(s)}}{\partial \boldsymbol\theta_{l,\nu\nu'}} &=\sigma_{\nu} \sigma_{\nu'}',\\
\frac{\partial f_l}{\partial \boldsymbol\theta_{l,\nu\nu'}} &=-\hat{i}\sum_{\mu,\mu'}\mathrm{Tr}(\sigma_{\mu} \sigma_{\mu'}' \sigma_{\nu} \sigma_{\nu'}'U_l^{(s)})\langle \phi_{l+1}^{\leftarrow}|\sigma_{\mu} \sigma_{\mu'}'|\phi_{l}^{\rightarrow}\rangle.
\end{align}

\begin{figure}
	\includegraphics[width=16cm]{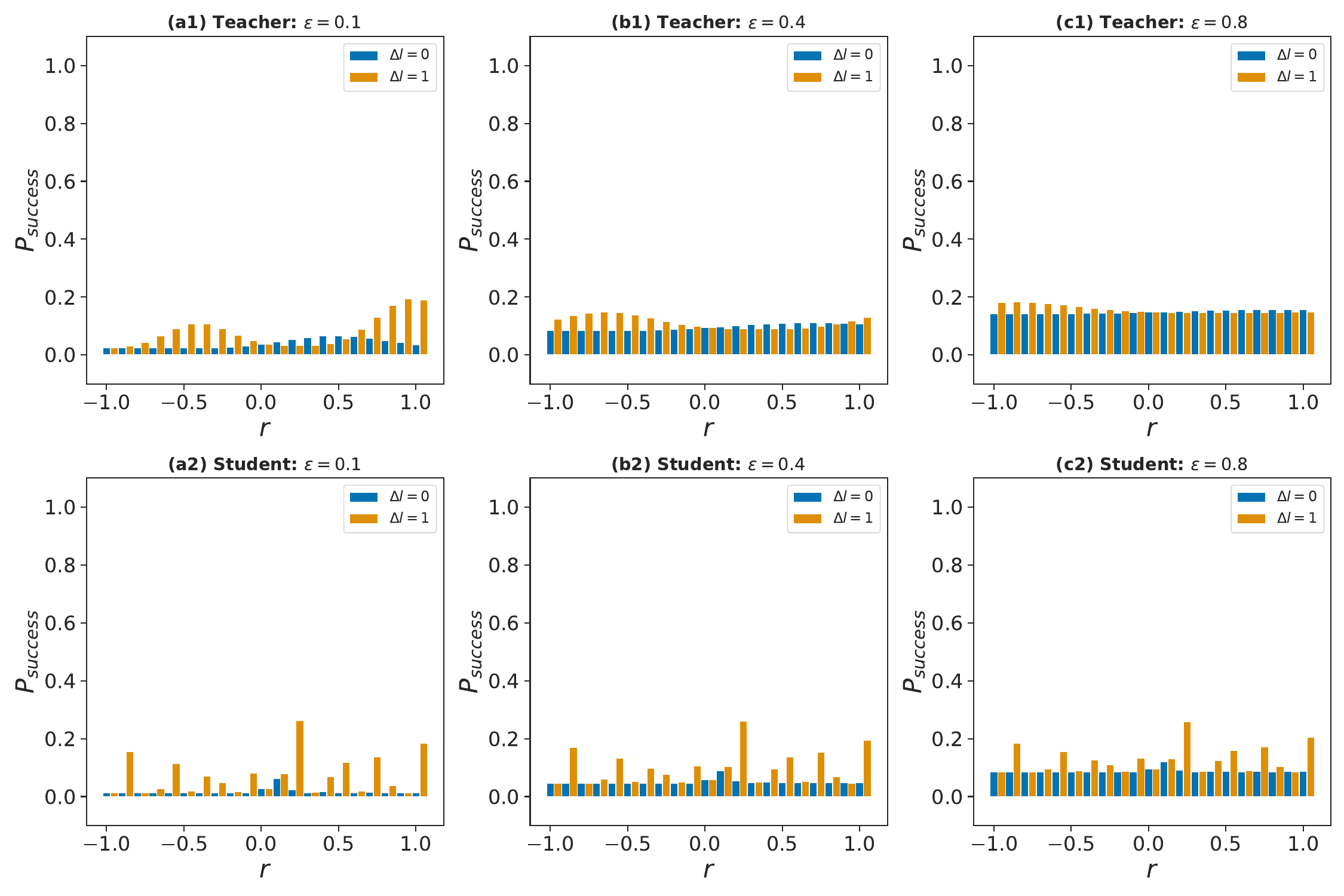} 
	\caption{Two qubits: Success probability under depolarizing noise. Here $L_t=10$ and $L_s=5$. ((a1), (b1), (c1)) For the teacher model. ((a2), (b2), (c2)) For the student model. The student results are obtained after $100$ GD iterations with learning rates $\eta=0.02$.}\label{fig-B1}
\end{figure}

\begin{figure}
	\includegraphics[width=16cm]{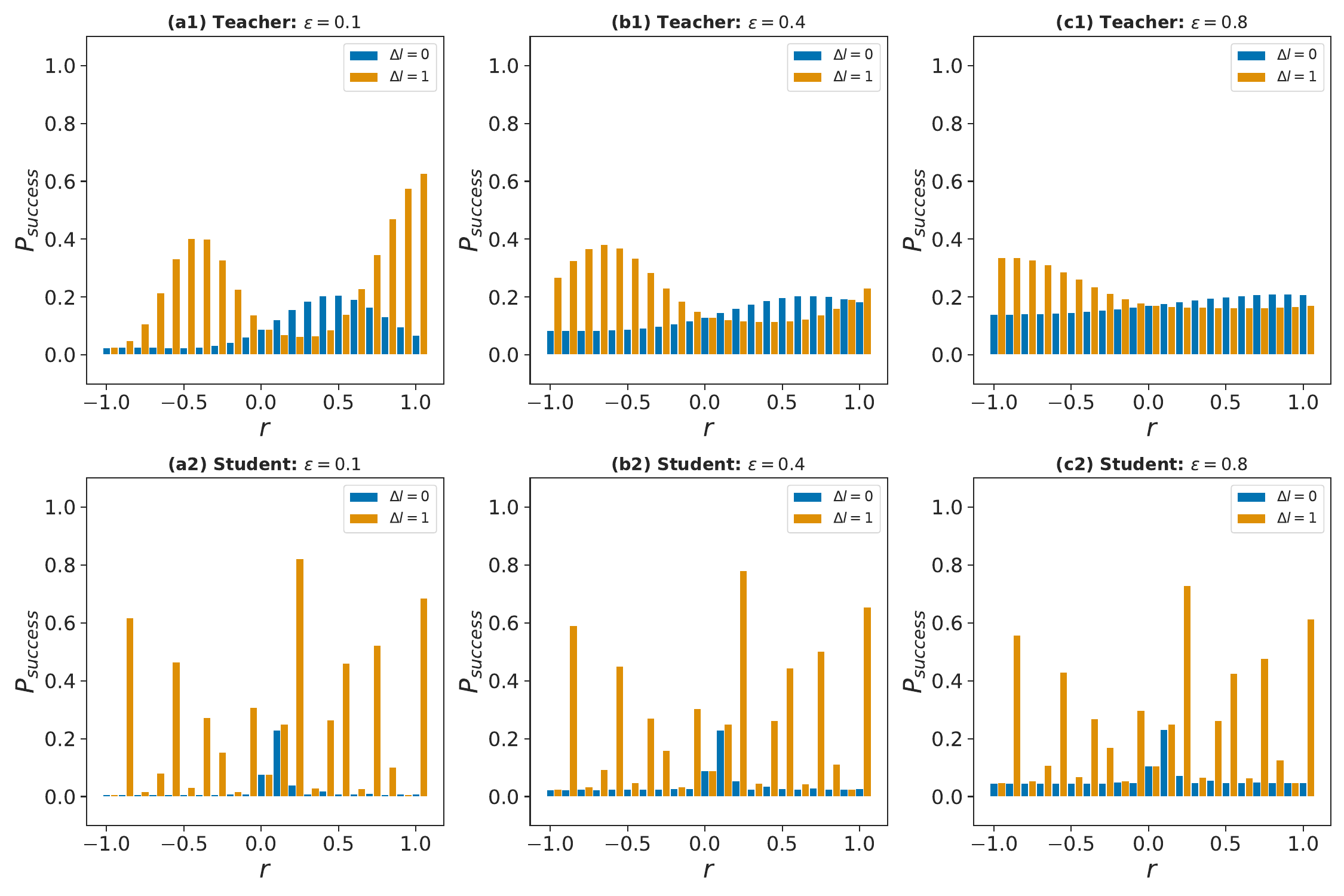} 
	\caption{Two qubits: Success probability under depolarizing noise. Here $L_t=20$ and $L_s=5$. ((a1), (b1), (c1)) For the teacher model. ((a2), (b2), (c2)) For the student model. The student results are obtained after $100$ GD iterations with learning rates $\eta=0.02$.}\label{fig-B2}
\end{figure}

\begin{figure}
	\includegraphics[width=16cm]{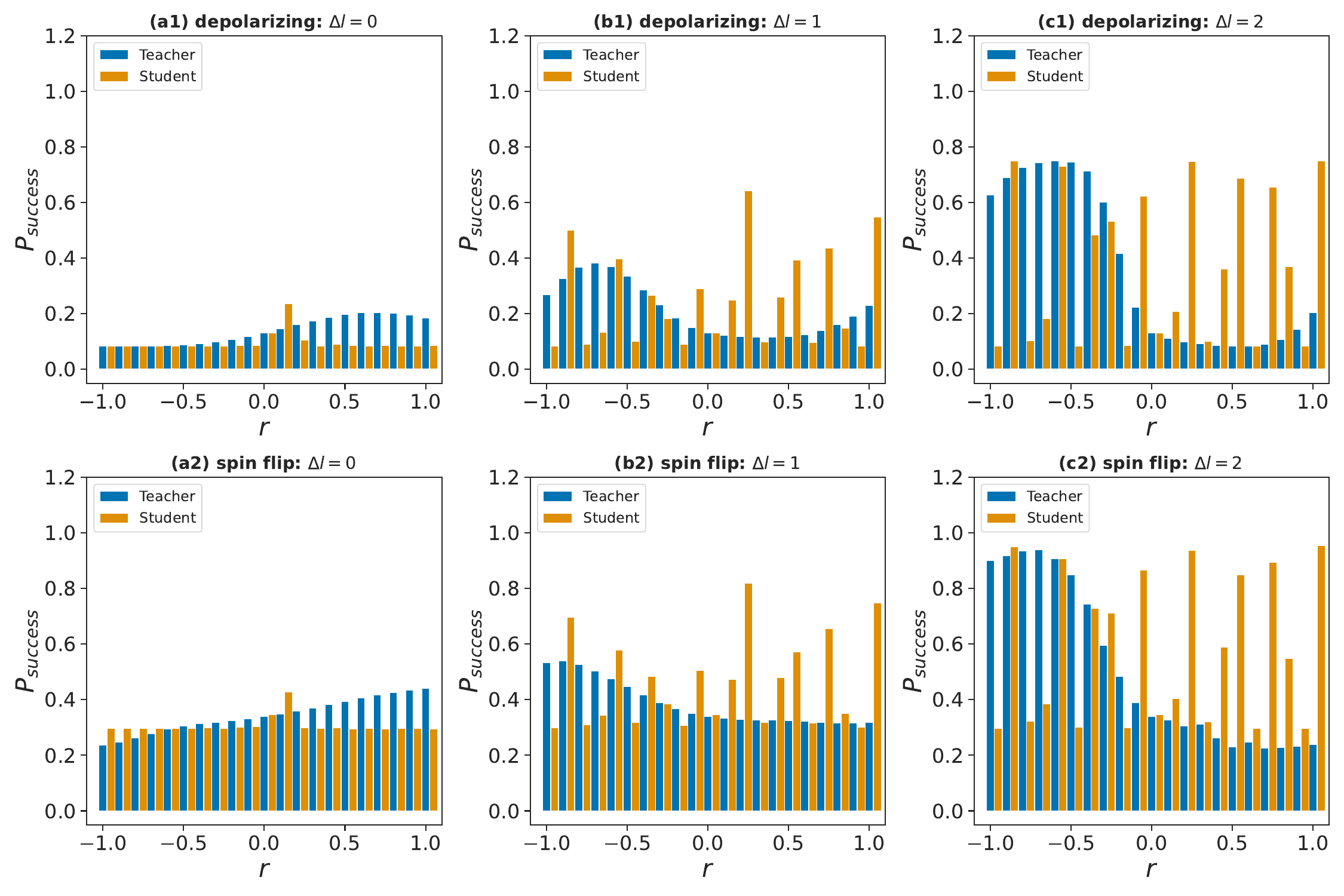} 
	\caption{Two qubits: Success probability under depolarizing and bit-flip noise. Here $L_t=L_s=20$, $\epsilon=0.4$. ((a1), (b1), (c1)) For depolarizing noise. ((a2), (b2), (c2)) For bit-flip noise. The student results are obtained after $100$ GD iterations with learning rates $\eta=0.02$.}\label{fig-B3}
\end{figure}

Working with the eigenstates $|n\rangle$ of the Hamiltonian $H_l^{(s)}|n\rangle=E_l(n)|n\rangle$, we get
\begin{align}
\mathrm{Tr}(\sigma_{\mu} \sigma_{\mu'}' \sigma_{\nu} \sigma_{\nu'}' U_l^{(s)}) &=\sum_n\langle n|\sigma_{\mu} \sigma_{\mu'}'\sigma_{\nu} \sigma_{\nu'}'|n\rangle e^{-\hat{i}E_l(n)}.
\end{align}

Finally, we obtain
\begin{align}
\Delta \theta_{l,\nu\nu'}=\eta \sum_n\sum_{\mu,\mu'}\mathrm{Im}\left(e^{-\hat{i}E_l(n)}\langle n|\sigma_{\mu} \sigma_{\mu'}'\sigma_{\nu} \sigma_{\nu'}'|n\rangle \langle \phi_{l+1}^{\leftarrow}|\sigma_{\mu} \sigma_{\mu'}'|\phi_{l}^{\rightarrow}\rangle \right).
\end{align}
In each iteration of the GD algorithm all parameters are updated once according to the above equation.

Figure \ref{fig-A0} (lower panels) shows how the learning error decreases in the GD algorithm with the parameters $\theta_{l,\mu\mu'}\in (-1,+1)$. The initial values are in $(-10^{-6},+10^{-6})$ and the learning rate $\eta=0.02$. Figures \ref{fig-B1} and \ref{fig-B2} display the success probabilities for different values of $r, \Delta l, \epsilon$ when $L_t=10, 20$ and $L_s=5$. In Fig. \ref{fig-B3} we see the teacher and student performances when $L_t=L_s=20$ in presence of depolarizing and bit-flip noise when $\epsilon=0.4$.

\end{document}